\documentclass[floats,floatfix,showpacs,amssymb,prd,twocolumn,superscriptaddress,nofootinbib,nolongbibliography,reprint]{revtex4-2}
\usepackage[utf8]{inputenc}
\usepackage{amsmath}
\usepackage{amssymb}
\usepackage{pifont}
\usepackage{bm}
\usepackage{enumitem}
\usepackage[acronym]{glossaries}
\usepackage{graphicx}
\usepackage[dvipsnames, usenames]{xcolor}
\definecolor{linkcolor}{rgb}{0.0,0.3,0.5}
\usepackage[unicode, colorlinks=true, linkcolor=linkcolor, citecolor=linkcolor, filecolor=linkcolor,urlcolor=linkcolor, pdfusetitle]{hyperref}
\usepackage[capitalise]{cleveref}  
\crefname{section}{Sec.}{Secs.}

\usepackage{mathtools}
\usepackage{multirow}
\usepackage{booktabs}
\usepackage{physics}
\usepackage{siunitx}
\usepackage{orcidlink}
\usepackage[normalem]{ulem}
\usepackage{appendix}

\makeglossaries
\newacronym{dm}{DM}{dark matter}
\newacronym{ce}{CE}{Cosmic Explorer}
\newacronym{et}{ET}{Einstein Telescope}
\newacronym{snr}{SNR}{signal-to-noise ratio}

\makeatletter
\renewcommand\onecolumngrid{%
\do@columngrid{one}{\@ne}%
\def\set@footnotewidth{\onecolumngrid}%
\def\footnoterule{\kern-6pt\hrule width 1.5in\kern6pt}%
}
\makeatother

\DeclareSIUnit\solarmass{\ensuremath{\mathrm{M}_\odot}}
\DeclareSIUnit\parsec{pc}
\DeclareSIUnit\gigaparsec{Gpc}
\DeclareSIUnit\year{yr}

\newcommand{\milan}{Dipartimento di Fisica ``G. Occhialini'', 
Universit\`a degli Studi di Milano-Bicocca, Piazza della Scienza 3, 20126 Milano, Italy}
\newcommand{\infn}{INFN, Sezione di Milano-Bicocca, 
Piazza della Scienza 3, 20126 Milano, Italy}

\newcommand{\GRAPPA}{Gravitation Astroparticle Physics Amsterdam (GRAPPA),\\ Institute for Theoretical Physics Amsterdam and Delta Institute for Theoretical Physics,\\ University of Amsterdam, Science Park 904, 1098 XH Amsterdam, The Netherlands}

\newcommand{\esa}{European Space Agency (ESA), European Space Research and Technology Centre (ESTEC), Keplerlaan 1, 2201 AZ Noordwijk, the Netherlands 
}

\begin{document}

\title{Sequential simulation-based inference for extreme mass ratio inspirals}

\author{Philippa S. Cole$\,$\orcidlink{0000-0001-6045-6358}}
\email{philippa.cole@unimib.it}
\affiliation{\milan}\affiliation{\infn}

\author{James Alvey}
\affiliation{Kavli Institute for Cosmology Cambridge, Madingley Road, Cambridge CB3 0HA, United Kingdom}
\affiliation{Institute of Astronomy, University of Cambridge, Madingley Road, Cambridge CB3 0HA, United Kingdom}

\author{Lorenzo Speri$\,$\orcidlink{0000-0002-5442-7267}}
\affiliation{\esa}

\author{\\Christoph Weniger}
\affiliation{\GRAPPA}

\author{Uddipta Bhardwaj}
\affiliation{\GRAPPA}

\author{Davide Gerosa$\,$\orcidlink{0000-0002-0933-3579}}
\affiliation{\milan}\affiliation{\infn}

\author{Gianfranco Bertone}
\affiliation{\GRAPPA}

\begin{abstract}
\noindent Extreme mass-ratio inspirals pose a difficult challenge in terms of both search and parameter estimation for upcoming space-based gravitational-wave detectors such as LISA. Their signals are long and of complex morphology, meaning they carry a large amount of information about their source, but their waveforms are expensive to compute and they occupy a vast and multi-modal parameter space. We explore how sequential simulation-based inference methods, specifically truncated marginal neural ratio estimation, could offer solutions to some of the challenges surrounding extreme-mass-ratio inspiral data analysis. We show that this method can efficiently narrow down the volume of the complex 11-dimensional search parameter space by a factor of $10^6-10^7$ and provide 1-dimensional marginal proposal distributions for non-spinning extreme-mass-ratio inspirals. We discuss the current limitations of this approach and place it in the broader context of a global strategy for future space-based gravitational-wave data analysis.
\end{abstract}

\maketitle

\section{Introduction}
\noindent Extreme-mass-ratio inspirals (EMRIs) are a key target for future space-based gravitational wave detectors such as the Laser Interferometer Space Antenna (LISA) \cite{LISA:2024hlh}, as well as TianQin \cite{TianQin:2015yph} and Taiji \cite{Hu:2017mde}. Defined as a binary black-hole system where the primary mass is much larger than the secondary, their interesting dynamics have a rich phenomenology that could offer new astrophysics and fundamental physics insights \cite{Babak:2017tow}. For example, EMRIs are ideal testbeds for beyond-General-Relativity effects \cite{Han:2018hby,Gupta:2022fbe,Speri:2024qak,Kejriwal:2023djc,Cardenas-Avendano:2024mqp}, environmental effects due to dark matter \cite{Eda:2013gg,Macedo:2013qea,Kavanagh:2020cfn,Khalvati:2024tzz,Vicente:2025gsg}, scalar fields \cite{Baumann:2021fkf,Barsanti:2022vvl,Zhang:2022rfr,Tomaselli:2024dbw}, or accretion disks \cite{Cole:2022yzw,Speri:2022upm,Duque:2024mfw,Copparoni:2025jhq}, as well as probing intermediate and supermassive black-hole formation mechanisms \cite{Volonteri:2010wz}. 

However, with great richness comes great complication. EMRIs are expected to have eccentric and inclined orbits and many thousands of harmonics contributing to the gravitational waveform, which makes computing the waveforms challenging. The validity of post-Newtonian methods for long duration waveforms do not extend down to such small mass ratios and numerical-relativity simulations of systems with such large differences in scales is computationally unfeasible, see Ref.~\cite{Wittek:2024gxn} for attempts to resolve this. Furthermore, these systems are expected to remain in the LISA band for months or years, meaning that many thousands or millions of orbits must be computed. Currently, adiabatic approximations~\cite{Hughes:2021exa}, effective-one-body approaches \cite{Buonanno:1998gg,Yunes:2010zj,Albertini:2023aol} and self-force theory \cite{Poisson:2011nh,Pound:2015tma,Wardell:2015kea,Barack:2018yvs,Pound:2019lzj,Pound:2021qin,Burke:2023lno} are relied upon, and it is an active area of research to beat down waveform systematics from these methods so that they are accurate enough to be used in gravitational-wave data analysis.

On the detector side, the challenges of extracting signals from realistic space-based gravitational-wave detector data are numerous \cite{Speri:2022kaq}. The noise is expected to be non-stationary and non-Gaussian, due to glitches, drifting noise levels in the components of the instruments, as well as having gaps due to both planned and unplanned downtime \cite{Dey:2021dem,Baghi:2021tfd,Littenberg:2023xpl,Spadaro:2023muy,Houba:2024tyn,Houba:2024ysm,Sauter:2025iey,Mao:2024jad,Castelli:2024sdb,Burke:2025bun}. Furthermore, there will be many overlapping signals in the data, including thousands of quasi-monochromatic galactic sources that will remain in band for the duration of the LISA mission \cite{Crowder:2004ca,Antonelli:2021vwg,Strub:2022upl,Niu:2024wdi}.

Searches and parameter-estimation algorithms for ground-based gravitational-wave detector data traditionally rely upon match-filtering and likelihood-based methods \cite{Chatziioannou:2024hju}. However, given the computational expense of computing a waveform template bank large enough for the search portion -- $\sim10^{40}$ templates will be required to perform a fully coherent matched filter search for year-long inspiral waveforms \cite{Gair:2004iv,Moore:2019pke} -- and the highly complex likelihood surface in the context of parameter estimation, these methods may not be currently optimised for space-based gravitational-wave data.

In this paper, we explore how simulation-based inference methods, which have been shown to be successful in other gravitational-wave applications \cite{Dax:2021tsq,Crisostomi:2023tle,Bhardwaj:2023xph,Raymond:2024xzj,Pacilio:2024qcq,Dax:2024mcn,Alvey:2023npw}, can help tackle the challenges of EMRI searches and parameter estimation. We build a pipeline to extract the parameters of Schwarzschild EMRIs in vacuum using Truncated Marginal Neural Ratio Estimation (TMNRE)~\cite{Miller:2021hys,Miller:2022shs}, assuming Gaussian noise and including the LISA antenna response. We show that this sequential method can be a very powerful tool for narrowing down the parameter space from initially wide priors, emulating a search strategy, as well as producing proposal distributions that encompass the true injected values. We note here that in the LISA context, the ``search" does not constitute isolating a short-duration signal from within a much longer data stream that is predominantly noise, as is the case for current ground-based gravitational wave detectors. The LISA search will involve extracting signals from multiple sources that are simultaneously present in the data stream as part of a global fit \cite{Katz:2024oqg}. So far, EMRIs have not been included in such a pipeline, but we imagine the search problem for EMRIs to comprise analysing data residuals after the removal of louder source classes such as massive black hole binaries to gain estimates on the parameters of any EMRI signals present. We therefore refer to the reduction in prior volume for realistic values of the parameters of the source as the search. Parameter estimation then corresponds to measuring the values of the system's parameters precisely having identified that a signal is present.

This paper is organised as follows: in \cref{sec:FEW}, we describe how we produce EMRI waveforms for this study; in \cref{sec:detector}, we briefly describe the noise and detector response that we employ; in \cref{sec:stateofart}, we review the state of the art for EMRI searches and parameter estimation;  in \cref{sec:tmnre}, we introduce our TMNRE approach; in \cref{sec:results},  we present our results; in \cref{sec:outlook}, we discuss future prospects.

\section{EMRI waveforms}\label{sec:FEW}

\noindent We model EMRI signals using the time domain implementation of the Fast EMRI Waveform (FEW) package~\cite{Katz:2021yft,fastfourier, Chua:2020stf,michael_l_katz_2023_8190418,Barack:2003fp,Chua:2017ujo,Chua:2015mua,Stein:2019buj,Fujita:2020zxe,Chua:2018woh}, which is GPU-accelerated. The six parameters that describe a Schwarzschild EMRI system in the source frame\footnote{Note that \texttt{FEW} waveform generator takes detector masses as input.} are the primary black-hole mass $m_1$, the secondary black-hole mass $m_2$, the initial semi-latus rectum $p_0$, the initial eccentricity of the system $e$, and the initial radial and azimuthal phases $\Phi_{r,0}$ and $\Phi_{\phi,0}$. Five extrinsic parameters are then required to describe the source in the detector frame (see \cref{sec:detector}), namely the polar angle of the direction of the orbital angular momentum unit vector $\theta_k$, the azimuthal angle of the direction of the orbital angular momentum unit vector $\phi_k$, the luminosity distance $d_L$, the polar sky location angle $\theta_S$, and the azimuthal sky location angle $\phi_S$.

\texttt{FEW} computes the phase and the amplitude of the gravitational waveform up to first order in gravitational self-force theory, i.e. the adiabatic approximation, which relies on expanding the metric of the binary in powers of the mass ratio $q=m_2/m_1$. The amplitudes of the modes are computed to lower accuracy, and are pre-computed for a subset of geodesic parameter sets, $\{p,e\}$ in the Schwarzschild case, and interpolated to cover the wider parameter space using a combination of reduced-order-model \cite{Field:2011mf} surrogates and deep-learning methods. The amplitude and the phase are combined for each mode, and the modes are summed over to produce an adiabatic waveform $h(t)=h_+(t) - ih_\times(t)$ in the time domain.

While in this paper we only consider Schwarzschild adiabatic (1st-order self-force) systems, we stress that the simulation module can be easily adapted to include more parameters and different waveform models, so we expect to be able to extend our results to the Kerr case (requiring three extra parameters: the dimensionless spin of the massive black hole $a$, the
inclination angle between the orbital angular momentum
unit vector and the spin vector $x_I$, and the initial polar
phase $\Phi_{\theta,0}$), and to include 2nd-order self-force implementations in the near future, see e.g. Ref.~\cite{Burke:2023lno}. We acknowledge however, that whilst including extra parameters or more complex waveforms in the simulator is simple, they also introduce additional degeneracies to the already vast and multi-modal parameter space. It is not immediately apparent how well the current network architecture will cope with these, and adaptations may be required to extract new features from the signals.

\section{LISA response and noise}\label{sec:detector}

\noindent LISA will consist of three satellites in heliocentric orbits, forming a nearly equilateral triangular constellation with arm lengths of approximately $2.5\times10^6\,{\rm km}$~\cite{LISA:2024hlh}. Each spacecraft will exchange laser beams with the other two, and measuring light propagation times in both directions results in six interferometric links. Incoming gravitational waves induce perturbations in these light-travel times, which can be extracted via time-delay interferometry (TDI), a technique that cancels otherwise overwhelming laser phase noise by constructing specific linear combinations of the delayed signals \cite{Tinto:2004wu,1999ApJ...527..814A,Tinto:2018kij,Tinto:2003vj,Muratore:2022nbh}.

The Michelson variables, $X, Y, Z$ have correlated noise properties, whereas linear combinations of these, the so-called $A, E$ and $T$ channels, are orthogonal and  uncorrelated in the approximation that the arm lengths are fixed and equal. In reality, both the ``breathing'' motion of the constellation, as well as orbital adjustments to combat drift, break this orthogonality. For simplicity and computational efficiency, we do not use the $T$ channel in our analysis, as it suppresses gravitational-wave power with respect to the other two, although see e.g. Ref.~\cite{Hartwig:2023pft} for more details. We feed our simulated time-domain signals into the \texttt{fastlisaresponse}~\cite{Katz:2022yqe} code, which is GPU-accelerated, to compute the projection onto the $A$ and $E$ TDI variables. We use first generation TDI variables, whilst making use of the European Space Agency's simulations of LISA orbits \cite{2021JAnSc..68..402M} to project the signals onto the LISA arms. In this ``hybrid'' set-up \cite{Katz:2022yqe}, the $A$ and $E$ channels are not orthogonal. This set-up is expected to induce biases of order $0.1\%$ on sky localisation parameters with respect to second-generation TDI variables for galactic binaries \cite{Katz:2022yqe}, and we note that it will not be sufficient to suppress laser noise down to the required level. We will implement second-generation TDI variables in future work now that compatible power spectral densities (PSDs) are more readily available.

In this work, we assume stationary, Gaussian noise according to the PSD as given in Eq.~(58) of Ref.~\cite{radler} and implemented in \texttt{pycbc} v.2.4.0~\cite{alex_nitz_2023_10013996}. We consider TDI 1.0 variables and account for the removal of confusion noise due to galactic white-dwarf binaries using the analytic fit implemented in Ref.~\cite{alex_nitz_2023_10013996}. This is a function of the duration that LISA has been taking data for, which we take to be 1 year, and decreases the amplitude of the effective PSD around the frequencies $\lesssim10^{-3}\,{\rm Hz}$.

\section{State of the art}\label{sec:stateofart}

\noindent Various strategies have been explored for EMRI searches and parameter estimation. On the search side, EMRIs span such a vast parameter space that fully coherent template bank methods become impractical, since their posterior distributions occupy only a tiny fraction of this space \cite{Gair:2004iv}. As a result, conventional grid-based approaches are infeasible, leading to the adoption of stochastic search techniques. However, the presence of numerous local maxima within the search landscape due to intricate, non-local parameter degeneracies in the signal space \cite{Chua:2021aah} implies that many combinations of parameters can produce waveforms that closely match the data. Distinguishing the true global maximum is a complex puzzle, as it can be located several $\sigma$ away from the local maxima, see \cite{Chua:2021aah} for more details.

Previous works have adopted several strategies to address these problems. Stochastic searches that use hybrid methods between genetic algorithms and MCMC \cite{Cornish:2008zd}, as well as methods built around the Metropolis–Hastings algorithm~\cite{Gair:2008zc,Babak:2009ua} have been explored as a means for using information about local maxima to try and identify the global maximum.

An alternative approach was proposed in Ref.~\cite{Chua:2022ssg} which used a veto likelihood to remove the local maxima, allowing only the global maximum to remain.
Other approaches used phenomenological waveforms to first identify the EMRI harmonics before carrying out a parameter search \cite{Wang:2012xh}. More recently, Ref.~\cite{Ye:2023lok} performed a three-stage semi-coherent search beginning with physical harmonic waveforms followed by phenomenological waveforms. Ref.~\cite{Yun:2023vwa} used deep-learning techniques on time-frequency domain waveforms obtained in both the analytic-kludge and effective-one-body-Teukolsky frameworks to search for the mass, spin, and eccentricity of the injected systems. Furthermore, Ref.~\cite{Zhang:2022xuq} used convolutional neural networks to search for analytic-kludge and augmented analytic-kludge EMRI waveforms with noise according to the expected sensitivities of TianQin \cite{TianQin:2015yph}, assuming the detector behaves like two Michelson interferometers (i.e. no TDI). Finally, Ref.~\cite{Badger:2024rld} used sparse dictionary learning to search for and reconstruct analytic-kludge EMRI waveforms with primary mass and eccentricity parameters allowed to vary over broad ranges but with all other parameters fixed.

\begin{table}[t]
\centering
\begin{tabular}{@{}p{2cm}p{3cm}@{}}
\hline
\textbf{Parameter} & \textbf{Prior Range} \\ \hline
$m_1/10^5\,[M_\odot]$ & $\mathcal{U}[4, 9]$ \\
$m_2\,[M_\odot]$ & $\mathcal{U}[20, 60]$ \\
$p_0\,[m_1]$ & $\mathcal{U}[8.5, 14.0]$ \\
$e_0$ & $\mathcal{U}[0.05, 0.60]$ \\
$\cos \theta_K$ & $\mathcal{U}[-1, 1]$ \\
$\phi_K$ & $\mathcal{U}[0, 2\pi]$ \\
$100d_L\,[\rm Gpc]$ & $\mathcal{U}[8.75, 50.0]$ \\
$\Phi_{\phi, 0}$ & $\mathcal{U}[-\pi, \pi]$ \\
$\Phi_{r, 0}$ & $\mathcal{U}[-\pi, \pi]$ \\
$\cos \theta_S$ & $\mathcal{U}[-1, 1]$ \\
$\phi_S$ & $\mathcal{U}[0, 2\pi]$ \\ \hline
\end{tabular}
\caption{Uniform prior ranges used for all parameter estimation results shown in this paper.}
\label{tab:priors}
\end{table}

On the parameter estimation side, likelihood-based approaches have been the most utilised. For example, Ref.~\cite{Ali:2012zz} recovered EMRI signals from the mock LISA data challenge 1.B \cite{Babak:2008aa} employing a parallel tempering MCMC algorithm on two-week-long signals with the low-frequency approximation of the LISA response. The GPU-accelerated MCMC pipeline presented in Ref.~\cite{Saltas:2023qec} performs parameter estimation on noise-free analytic-kludge waveforms and a simplified response function. Works that utilise the \texttt{eryn} MCMC sampler \cite{Karnesis:2023ras}, which we use for comparison in \cref{sec:results}, include Refs.~\cite{fastfourier,Katz:2021yft}, which reconstruct a subset of parameters of noise-free frequency and time domain generated \texttt{FEW} signals, without including the LISA response; Ref.~\cite{Kejriwal:2025upp,Copparoni:2025jhq,Speri:2022upm}, which includes environmental effects; Ref.~\cite{Speri:2024qak} which investigates EMRI parameter estimation in beyond general relativity theories; Ref~\cite{Burke:2023lno} which includes post-adiabatic effects, and; Ref.~\cite{Toscani:2023gdf} which explores strong lensing.

More recently, simulation-based inference (SBI) methods have been investigated as an alternative to likelihood-based approaches. Normalizing flows are used in Ref.~\cite{2024arXiv240907957L} to reconstruct the parameters of very similar signals to those we use in this work, namely Schwarzschild \texttt{FEW} waveforms with the LISA response implemented as in Ref.~\cite{Katz:2022yqe}, however without noise.

We will now describe our investigation of using a sequential simulation-based inference method for EMRI search and parameter estimation. To the best of our knowledge, this is the first time that sequential simulation-based methods have been used to reconstruct all 11 parameters of a Schwarzschild EMRI waveform computed with \texttt{FEW}, including the LISA response function and Gaussian, stationary noise according to the LISA PSD. Crucially, we also start our analysis from broad priors, see \cref{tab:priors}, and in no sense do we initialise our pipeline close to the injected values that we want to infer. For example, Ref.~\cite{Babak:2009ua} searched the Mock LISA Data Challenge 1.B dataset \cite{Babak:2008aa} for low-mass EMRIs using the following priors (we only report parameters common to our analysis): $m_1 \in \mathcal{U}[0.95, 1.05] \times 10^6\,{\rm M_\odot}$, $m_2 \in \mathcal{U}[9.5, 10.5]\,{\rm M_\odot}$, and eccentricity at plunge (note we use instead initial eccentricity) $e_{pl} \in \mathcal{U}[0.15,0.25]$. Ref.~\cite{Cornish:2008zd} use $m_1 \in \mathcal{U}[4.8, 5.4] \times 10^6\,{\rm M_\odot}$, $m_2 \in \mathcal{U}[9.4, 11.4]\,{\rm M_\odot}$, and eccentricity $e \in \mathcal{U}[0.195,0.23]$. The most similar work to ours that uses simulation-based inference \cite{liang2024} used $m_1 \in \mathcal{U}[0.95, 1.1] \times 10^6\,{\rm M_\odot}$, $m_2 \in \mathcal{U}[9, 11]\,{\rm M_\odot}$, $e_0 \in \mathcal{U}[0.1,0.3]$, $p_0/m_1 \in \mathcal{U}[9.1,9.3]$ and the full domains for the four angular parameters. It is unclear how they treat the initial phases. The broadest priors implemented in the literature that we refer to in this section are those of Ref.~\cite{Yun:2023vwa}, $m_1 \in \mathcal{U}[10^5, 10^8]\,{\rm M_\odot}$, $m_2 \in \mathcal{U}[10,100]\,{\rm M_\odot}$, $e \in \mathcal{U}[0.005,0.6]$, $p_0/m_1 \in \mathcal{U}[10, 12]$ and Ref.~\cite{Ye:2023lok}: $m_1 \in \mathcal{U}\ln[10^5, 10^7]\,{\rm M_\odot}$, $m_2 \in \mathcal{U}[5,15]\,{\rm M_\odot}$, $e \in \mathcal{U}[0,0.7]$,
with a fixed luminosity distance of $1\,{\rm Gpc}$. Both of these references use full domains for the angular parameters.

\section{Truncated Marginal Neural Ratio Estimation }\label{sec:tmnre}

\noindent We use TMNRE \cite{Miller:2021hys,Miller:2022shs} in order to run parameter estimation with wide priors on EMRI systems. TMNRE is one implementation of sequential simulation-based inference, see Ref.~\cite{sbireview} for a review of this class of methods. Specifically, we developed a version of the \texttt{peregrine} software~\cite{Bhardwaj:2023xph} suitable for EMRIs. The \texttt{peregrine} code was initially developed to be an optimised implementation of the TMNRE scheme for gravitational-wave signals at ground-based interferometers and was shown to perform well in the case of overlapping events \cite{Alvey:2023naa}. Other use cases of the same scheme in the context of gravitational waves include \texttt{saqqara}~\cite{Alvey:2023npw,Alvey:2024uoc} for inferring the noise properties of LISA from stochastic background observations. See also e.g. Refs.~\cite{FrancoAbellan:2024tbj,Alvey:2023pkx,AnauMontel:2022ppb,Cole:2021gwr} for other astrophysical and cosmological use cases.

At the basic level, TMNRE trains a neural network to approximate the likelihood-to-evidence ratio, which, when re-weighted by prior samples, yields an estimate of the posterior distributions over the parameters of interest. Unlike direct joint posterior estimation (via likelihood- or simulation-based approaches), TMNRE learns lower (e.g. 1-d) dimensional marginal ratios directly. This has been shown to have the potential to significantly reduce the number of required waveform evaluations compared to full joint posterior estimation followed by marginalisation~\cite{Bhardwaj:2023xph}.

From a machine-learning perspective, the training objective for TMNRE is to minimise the binary cross-entropy loss~\cite{Miller:2021hys, Miller:2022shs}:
\begin{align}
      \mathcal{L}[\hat{\rho}_{k, \phi}] = &-
\int \big\{ p(x, \theta_k) \ln \sigma(\hat{\rho}_{k,\phi}(x, \theta_k)) 
\notag \\ & +\, p(x) p(\theta_k) \ln \left [1 - \sigma(\hat{\rho}_{k,\phi}(x, \theta_k)) \right] \big\} \, \dd x  \, \dd \theta_k\,.
\label{eq:loss}
\end{align}
Here, $x$ denotes the simulated data (including both signal and noise), $\theta_k$ represents some subset of the full parameter vector $\theta$, $\sigma$ is the sigmoid function, and $\phi$ denotes the parameters (weights and biases) of the neural network. In addition, $\hat{\rho}_{k,\phi}(x, \theta_k)$ represents the parameterised network trained to estimate the marginal posterior corresponding to the parameter set $\theta_k$. In contrast to density estimation, this formulation treats the problem as a simple binary classification task, requiring the network to distinguish between samples drawn jointly $x, \theta_k \sim p(x,\theta_k)$ versus those drawn marginally $x, \theta_k \sim p(x)p(\theta)$, where here $p(x)$ is the marginal distribution of the data $x$, i.e. the joint distribution $p(x, \theta)$ marginalised over $\theta$, and $p(\theta_k)$ is the prior over the marginal parameters of interest. The main motivation for this is that it can be shown analytically (by taking a functional derivative with respect to $\hat{\rho}_{k,\phi}$) that the optimal classifier $\rho^\star_{k,\phi}$ is precisely $\rho^\star_{k,\phi} = \ln [p(x | \theta_k) / p(x)]$~\cite{Miller:2021hys}. In practice, $x, \theta_k \sim p(x,\theta_k)$ is obtained by drawing a parameter sample $\theta \sim p(\theta)$, simulating the data $x$ that corresponds to those parameter values, and then only keeping the parts of $\theta_k \subset \theta$ that one wants to perform inference on. The samples $x, \theta_k \sim p(x)p(\theta_k)$ are obtained by the same procedure, followed by re-sampling the parameters $\theta_k$ so that the data was not generated according to the resampled parameter set. From a technical perspective, we minimise the loss using \texttt{AdamW} optimiser \cite{2017arXiv171105101L}, and monitor the training and validation loss (evaluated at the end of each epoch on an independent dataset) throughout with an early stopping criteria imposed to avoid overfitting.

Simulation and training proceed in iterative rounds, with sequential truncation of the prior range~\cite{Miller:2021hys,Miller:2022shs}. In each round, new simulations are generated from the current truncated prior, the network is trained, and updated posteriors (or proposal distributions) are estimated. We refer to the estimated posteriors as proposal distributions in the sense that they could inform the initial conditions of a subsequent parameter estimation pipeline, and are not to be confused with proposal distributions in the sequential SBI context. If the value of the likelihood-to-evidence ratio for a given parameter value is less than a predefined threshold (nominally $10^{-5}$), the prior is truncated accordingly, and the next round begins with a refined prior. This process continues until convergence, typically defined as the point where further truncation is negligible due to significant weight persisting at the edges of the final prior. In the current setup \cite{Bhardwaj:2023xph}, it is only possible to truncate from the outer edges of the marginal prior which, as we will see later, is sub-optimal for parameters which exhibit multi-modality. This is more of a technical hurdle than a fundamental one, however, and other options already exist to generate targeted training data including autoregressive ratio estimation algorithms~\cite{AnauMontel:2023stj} or active learning/truncation techniques in the context of density estimation \cite{Alsing:2019xrx,2022arXiv221004815D} which could be useful in this context.

We show the time-domain data to the network in 3 channels: the time array, the $A$-channel signal + noise, and the $E$-channel signal + noise. First it is fed through a U-Net, then the data is fed through a linear compression network which results in a vector of 16 features per parameter for each channel. We retained the structure of both the U-Net and the linear compression network from Ref.~\cite{Bhardwaj:2023xph}, and manually optimised the dimensions of the layers in both. Finally this feature vector is given to the logratio $(\ln [p(x | \theta_k) / p(x)])$ estimator which is a residual network \cite{He:2015wrn} that performs the marginal classfication as defined in \texttt{swyft} \cite{Miller:2021hys,Miller:2022shs}. For all rounds, the training and validation batch sizes are set to 128, with 90\% of the simulation budget allocated for training and 10\% for validation. We use a learning rate of $10^{-4}$ and employ an early stopping criterion of seven epochs, such that training halts if the validation loss does not decrease for seven consecutive epochs. At this point, the network weights revert to the state corresponding to the minimum validation loss, and proposal distributions are computed from this configuration. To mitigate noise-induced biases, we apply a ``noise shuffling" strategy -- each epoch, the noise realisations shown to the network are shuffled within each batch. This has the benefit that the network is highly unlikely to be shown the same combinations of signal plus noise, without having to generate new noise realisations every epoch. This prevents the network from learning features of the noise as characteristics of the signals which would lead to overfitting or biased parameter inference.

\section{Results}\label{sec:results}

\begin{figure*}
    \centering
    \includegraphics[trim=1.3cm 0.1cm 1.6cm 1cm, clip,width=\textwidth]{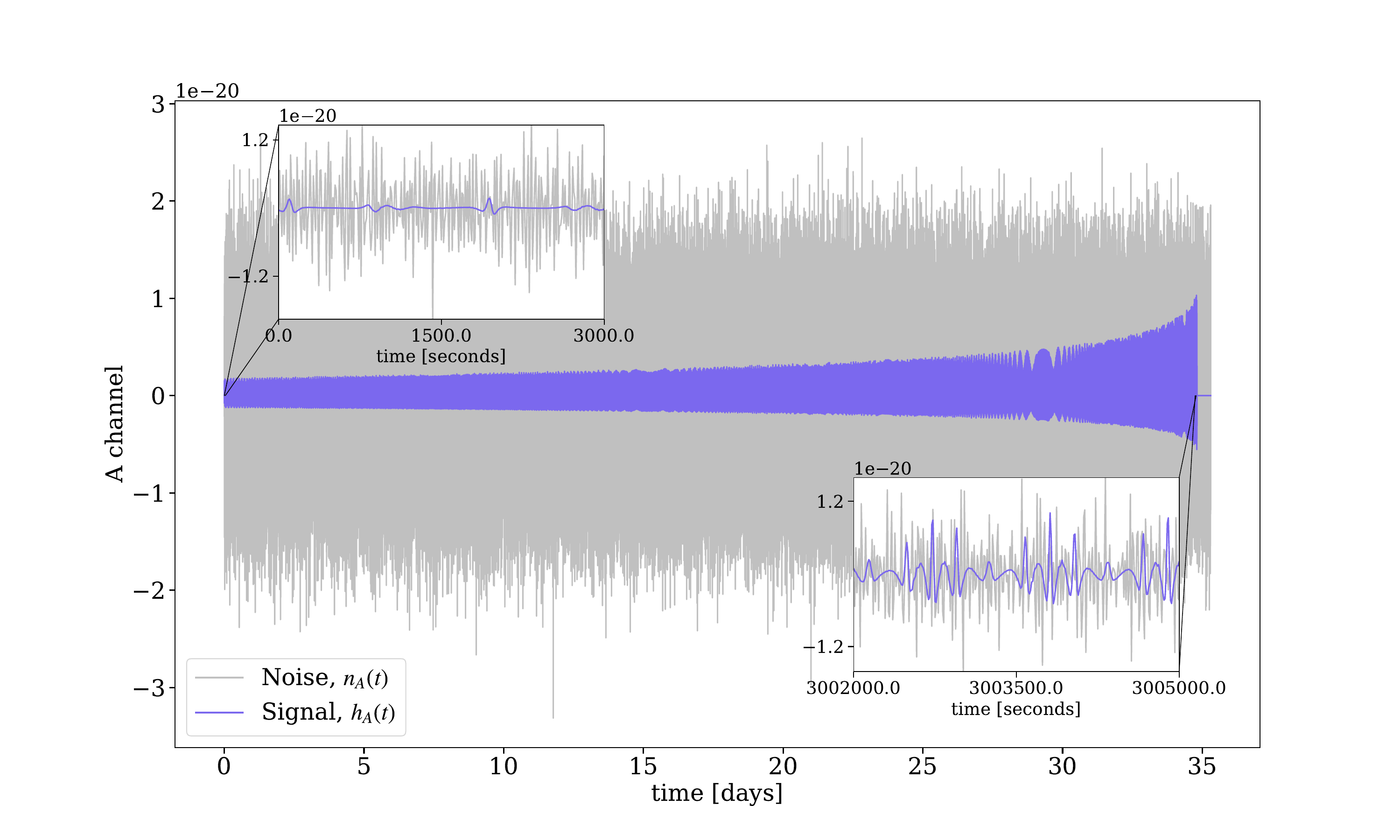}

    \caption{Injected time-domain signal (in purple) and noise (in grey) in the TDI $A$-channel for the fiducial system analysed in this work which has ${\rm SNR}\simeq640$. The two insets show zoom-in regions of the signal and noise at the beginning of the segment on the left and close to plunge on the right, with durations of 3000 seconds each.}
    \label{fig:signal}
\end{figure*}

\begin{figure*}
    \centering
    \includegraphics[width=\textwidth]{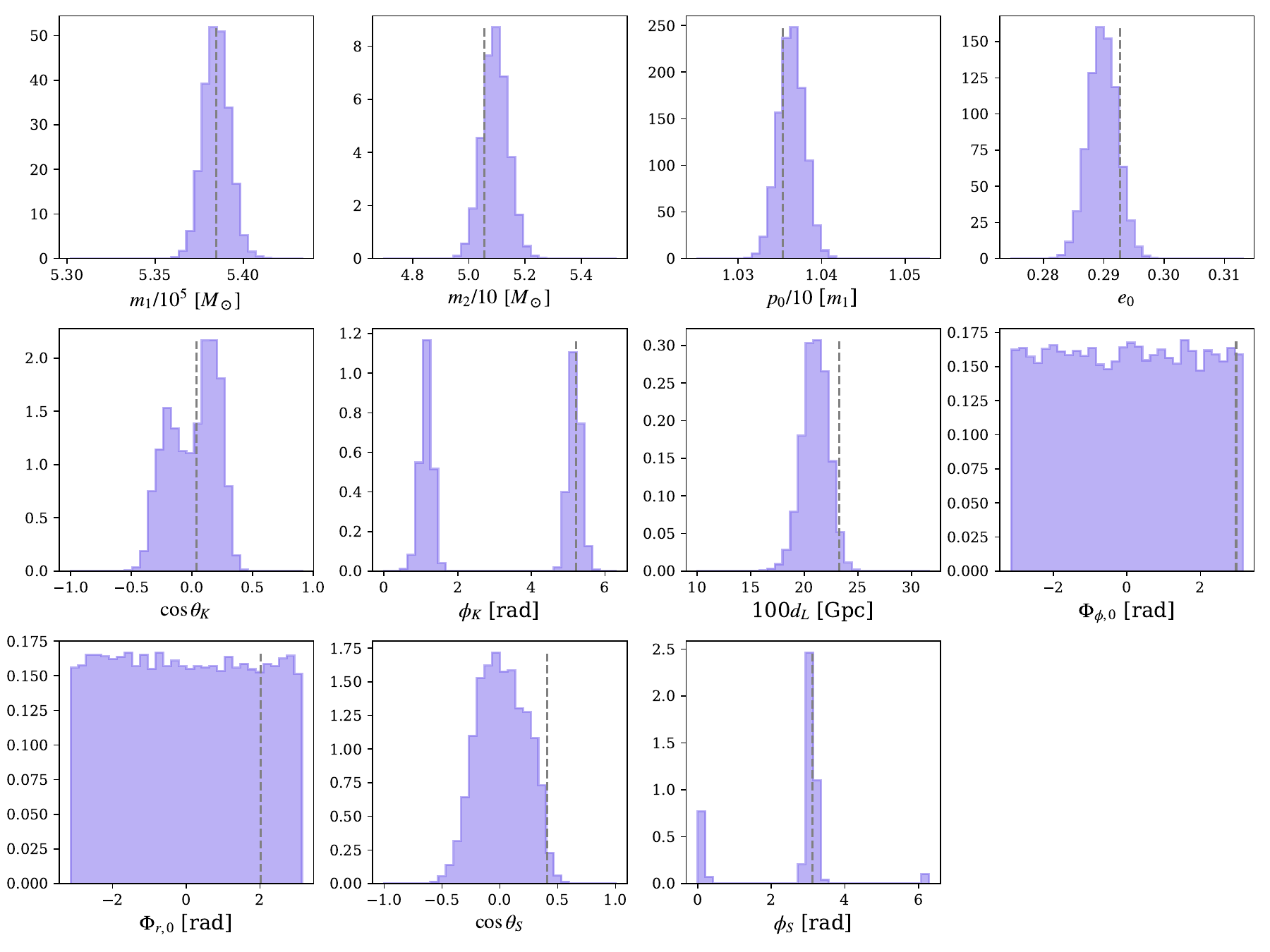}

    \caption{Marginal proposal distributions, i.e. high probability regions identified by the neural network, for the parameters of a non-spinning EMRI in eccentric orbits.  The injected signal considered is indicated by the grey dashed lines, and has ${\rm SNR \sim 640}$ with masses $m_1 = 5.385\times10^5\,{\rm M_\odot}$ and $m_2=50.55\,{\rm M_\odot}$, initial semi-latus rectum $p_0=10.35\,m_1$, initial eccentricity $e=0.2927$, cosine of the polar angle in the detector frame $\cos{\theta_K} = 0.0384$, azimuthal angle in the detector frame $\phi_K = 5.212$,  luminosity distance to the source $d_L = 232.8\,{\rm Mpc}$,  initial azimuthal phase $\Phi_{\phi} = 2.966$, initial radial phase $\Phi_{r} = 2.014$, cosine of the polar sky location coordinate $\cos{\theta_S = 0.4107}$, and azimuthal sky location coordinate $\phi_S = 3.124$. The signal is produced and analysed in the time domain, with a duration of $3.15\times10^6\,{\rm s}$, a sampling cadence of $\mathrm{d}t=10\,{\rm s}$ and is projected onto the $A$ and $E$ TDI channels. Noise according to the analytical LISA PSD is added. We ran TMNRE for six rounds, with $1.5\times 10^5$ simulations in each round.}
    \label{fig:PE_truncate}
\end{figure*}

\noindent We now present parameter estimation results for EMRI waveform injections whose parameters are sampled from within the prior ranges shown in \cref{tab:priors}. Signals are produced in the time domain, with a duration of $3.15\times10^6$ seconds (one tenth of a year) and a sampling cadence  $\mathrm{d}t=10\,{\rm s}$. This value of $\mathrm{d}t$ is small enough to avoid aliasing for the values of $m_1$ we consider here \cite{fastfourier}. We choose a mode content threshold of $\epsilon=10^{-5}$ as defined in \texttt{FEW}, meaning that if a particular mode contributes more than a fraction of $10^{-5}$ of the total power emitted by all modes then it is included in the summation when the waveform is evaluated. For the fiducial system we consider here, 581 modes are included.

We use wide priors within the domain of validity of \texttt{FEW}, in order to understand the potential of this algorithm for both searches (i.e. narrowing down the parameter space) and parameter estimation (subsequently measuring the parameters precisely), cf. Table~\ref{tab:priors}. The exceptions to this are the initial semi-latus recti, for which we choose the prior such that the simulations capture the final part of the merger, i.e. not time segments that are much earlier in the inspiral. Depending on the parameter set, some systems will merge just within the data segment, while others will not quite reach merger. We find (as expected) that the measurement precision on a given injection degrades considerably if the simulations only capture portions of the signals that are long before merger.

We are also limited by the memory budget on the HPC cluster that we use for this analysis; specifically, we run on an exclusive node with total 494 GB of memory, utilising a single NVIDIA A100 GPU with 64 GB of memory. With the current iteration of the algorithm, increasing the signal duration to be longer than 1/10 of year would require different hardware. In order to accumulate a signal-to-noise ratio that is sufficiently large for parameters to be recovered successfully, sources must be relatively closeby. The prior range on $d_L$ varies from the ``closest possible'' distance of an EMRI system, thought to be at $d_L \sim 87.5\,{\rm Mpc}$ or $z\sim0.02$~\cite{Gair:2008bx} up to $d_L=0.5\,{\rm Gpc}$ or $z\sim0.1$. We note that LISA is expected to observe EMRIs out to $z\sim2$ \cite{Mapelli_2012}, and probing larger distances requires analysing longer-duration signals, which in turn necessitate more sophisticated down-sampling and/or data-reduction schemes. See e.g. Ref.~\cite{Dax:2024mcn} for a method for simplifying and compressing binary neutron star gravitational wave data that utilises approximate information about the source. Innovations in this direction would alleviate the hardware constraints discussed above, and enable more information about the source to be extracted from the time evolution of the signal over a larger number of orbits. We expect this to improve the precision to which this pipeline is able to measure realistic EMRI parameters with lower SNRs, although it remains to be tested how the trade-off between longer signals (which narrow the posterior) and lower SNRs (which broaden the posterior and increase the number of secondaries) will be coped with by the sequential methods we investigate here, and we leave a full investigation to future work.

\subsection{Marginal proposal distributions}

\noindent For our first application, we consider an EMRI signal with the following injected parameters: primary mass $m_1 = 5.385\times10^5\,{\rm M_\odot}$, secondary mass $m_2=50.55\,{\rm M_\odot}$, initial semi-latus rectum $p_0=10.35\,m_1$, initial eccentricity $e=0.2927$,  cosine of the polar angle in the detector frame $\cos{\theta_K} = 0.0384$,  azimuthal angle in the detector frame $\phi_K = 5.212$,  luminosity distance to the source $d_L = 232.8\,{\rm Mpc}$, initial azimuthal phase $\Phi_{\phi} = 2.966$, initial radial phase $\Phi_{r} = 2.014$, cosine of the polar sky location coordinate $\cos{\theta_S = 0.4107}$, and azimuthal sky location coordinate $\phi_S = 3.124$. The system has ${\rm SNR}\simeq640$ and the duration of the inspiral up to merger is $3.1\times10^6\,{\rm s}$, with respect to the $3.15\times10^6\,{\rm s}$ observation window. The injected signal and noise in the $A$-channel are shown in \cref{fig:signal}. 
 
We show in \cref{fig:PE_truncate} the estimated 1D marginal proposal distributions produced with six rounds of sequential TMNRE. We note that these are the estimated posterior distributions output from the TMNRE pipeline, which we call proposal distributions as we expect them to be most useful in terms of reducing the prior volume for subsequent parameter estimation pipelines. We use 150000 simulations in each round and find that the algorithm converges after six rounds. We comment that the reduction in prior volume is not sufficient for a subsequent MCMC step if uniform priors are assumed, as discussed in \cref{sec:comparisons}, with the caveats that the MCMC settings were not optimised there and limited to 150000 waveform evaluations in order to compare fairly with the SBI set-up. However, we posit that the best-fit values of the TMNRE proposal distributions could be used as the initial state for an MCMC run, and we leave this exploration of a hybrid method for future work. 
 
The reconstructed proposal distributions are informative for all of the parameters except for the initial phases $\Phi_{\phi,0}$ and $\Phi_{r,0}$, where the algorithm returns the underlying uniform priors. This is likely due to the fact that we are resolving only the frequency evolution of the harmonics and not their relative phases. This also happens when secondary likelihood peaks are found due to degeneracies, as shown in Ref.~\cite{Chua:2021aah}. The initial phases $\Phi_{\phi,0}$ and $\Phi_{r,0}$ enter each harmonic $(m,n)$ as $m\Phi_{\phi,0} +n \Phi_{r,0}$.

In terms of precision, the relative half-widths of the proposal distributions for the intrinsic parameters, corresponding to the $2\sigma$ credible intervals, are $0.3\%$ for $m_1$, $2\%$ for $m_2$, $0.3\%$ for $p_0$ and $2\%$ for $e_0$. We comment on the comparison with the relative half-widths of typical MCMC posterior distributions in \cref{sec:comparisons}. We present various diagnostics of convergence in \cref{app:pp}, including $p-p$ plots and training/validation losses. Each round in this analysis took an average of 14 hours on 1 GPU, with $\sim$3 hours devoted to simulations and the remainder to training the network. Inference on a single observation (inside the truncated prior region) then takes a matter of seconds.

We note here that we also conducted this analysis in the frequency domain by taking a fast Fourier transform of the signal plus noise, and found comparable results.

\subsection{Comparisons}\label{sec:comparisons}

\noindent We now compare our results with those achievable with traditional likelihood-based methods in order to understand how simulation-based inference strategies could be useful in both the EMRI search and parameter-estimation space, noting that the pipeline attempts to do both sequentially.

Firstly, we comment on the fact that the TMNRE proposal distributions are broader than some of the posteriors obtained with MCMC methods in the EMRI literature, see e.g. Refs.~\cite{fastfourier,Gupta_2022,liang2024}. Very narrow posteriors, with relative half-widths of $10^{-3}-10^{-2}\%$ on the intrinsic parameters $\{m_1, m_2, p_0,e_0\}$ \cite{liang2024}, are achievable in reasonable computing time with MCMC methods if either narrow priors around the true values are used or the chains are initialised close to the true value. This allows the region of high posterior density to be identified immediately and avoids the complicated multi-modal distribution that must otherwise be explored. In this work, since we have used wide priors and a sequential truncation method that only bounds from the upper and lower limit of the prior, the extreme multi-modality ubiquitous with EMRI systems is one aspect of the problem which could be impacting the precision we can achieve. This is because the parameter space where there is zero weight in the proposal distribution in between two large peaks is not removed at the truncation step. We leave to future work a full investigation of a more sophisticated truncation method that would combat this multi-modality, as well as data-compression strategies that could better capture degeneracies between parameters and allow for the analysis of longer duration signals that exhibit more orbits. Improvements in these directions will help to achieve narrower proposal distributions that get closer to the true posteriors. 

\begin{figure*}
    \centering
    \includegraphics[trim=1.9cm 0.0cm 1.9cm 0.0cm, clip,width=\textwidth]{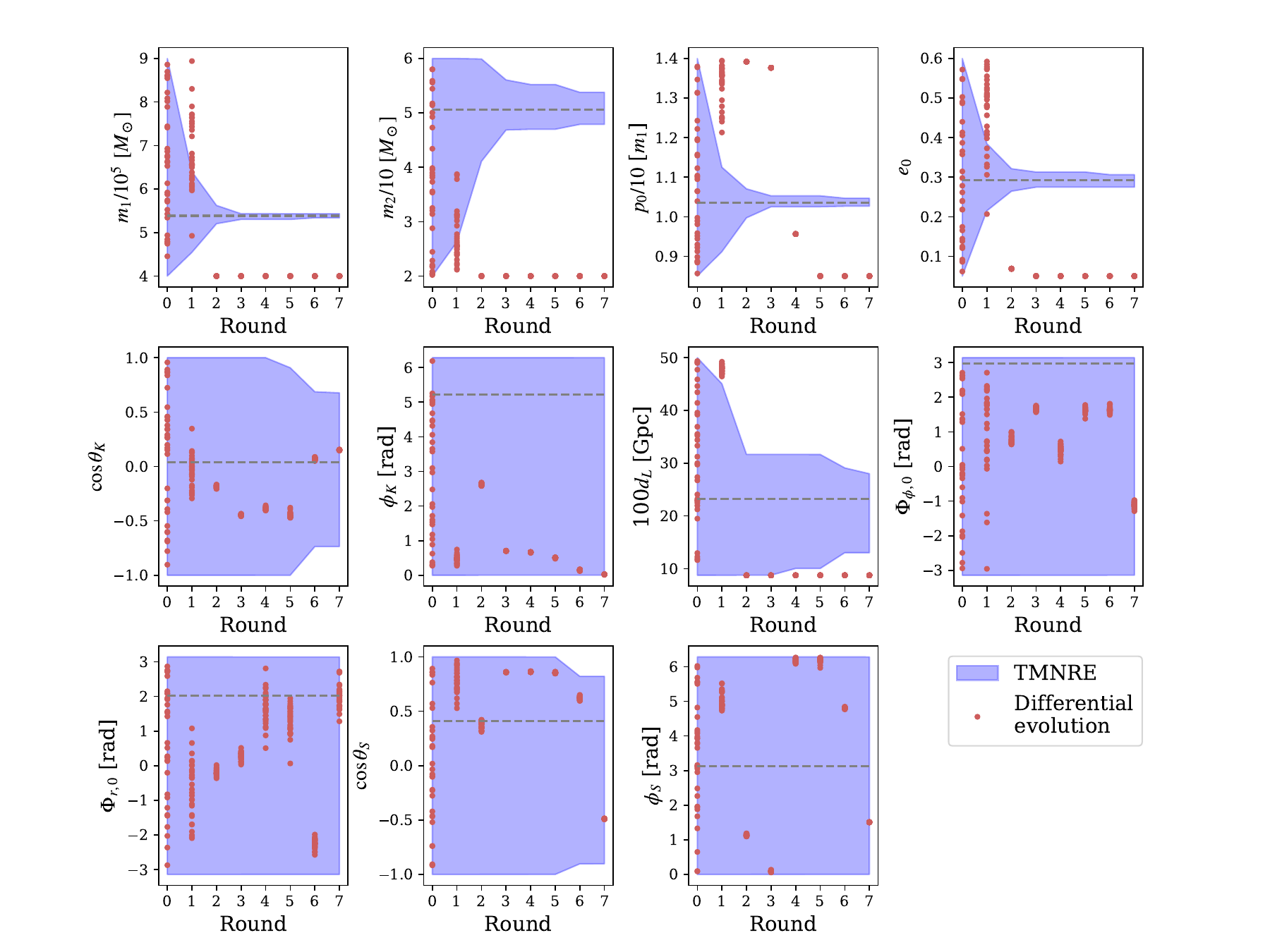}

    \caption{Comparison of TMNRE and differential evolution for narrowing down the parameter space over 7 sequential rounds. Each round of TMNRE uses 150000 simulations and the purple shaded region is defined by the upper and lower bounds that are placed on the parameter space at the end of each round, to be used for the following round's prior. The value of the likelihood-to-evidence ratio at parameter values outside of the shaded region is less than $10^{-5}$. For each round, the red points mark the solutions found by the differential-evolution algorithm for 32 population members. We limit the number of function evaluations in each round to be a maximum of 150000. The true injected values shown by the grey dashed lines are the same as for the system analysed in \cref{fig:PE_truncate}.}
    \label{fig:bound_compare}
\end{figure*}

Note, however, that assuming wide priors as done here brings us significantly closer to a real data-analysis scenario compared to previous EMRI studies. 
We demonstrate the success of this pipeline for narrowing down the prior volume drastically and its possible use as part of a search strategy, as well as its performance with respect to an MCMC that is not initiated with narrow priors and/or close to the true injected values.

One strategy for narrowing down the parameter space is to identify the parameter solution vector that best maximises the log-likelihood function before running, e.g., MCMC. This can be done with an multivariate optimisation algorithm such as differential evolution \cite{storn1997differential}, which is stochastic in nature and useful for searching large parameter spaces. We run differential evolution as implemented in \texttt{scipy} \cite{2020NatMe..17..261V} with 32 individuals drawn from the prior for 7 rounds, allowing a maximum of 150000 likelihood evaluations in each round (to match like-for-like our simulation budget). In \cref{fig:bound_compare}, we show the bounding on each parameter after sequential rounds of simulation and training with TMNRE versus the parameter solutions found by minimising the negative log-likelihood with differential evolution. TMNRE performs very effectively, especially for $m_1$, $m_2$, $p_0$, $e$ and $d_L$. Note that its performance for the angular parameters which exhibit multi-modality in the proposal distributions is not well expressed in \cref{fig:bound_compare} because it does not show the low-probability regions in-between large peaks. Note that optimising over a different statistic as opposed to the likelihood with differential evolution has been shown to be more effective in Ref.~\cite{strub2025searchingextrememassratio}, which appeared after the submission of this manuscript. 

We can quantify the reduction in the volume of the parameter space by comparing the initial 11-dimensional prior volume $V_{i}$ with the remaining volume after 6 rounds of TMNRE $V_{f}$. The ratio $V_i/V_f$ is $7 \times 10^5$, meaning we have narrowed down the parameter space volume by almost a factor of a million. Differential evolution, meanwhile, is not able to identify the correct regions of the parameter space that contain the true values, likely getting stuck in local minima of the negative log-likelihood rather than finding the true global minimum. These results demonstrate a very promising use case for sequential SBI as part of a search strategy for EMRIs. We show additional examples of narrowing down the parameter space for the intrinsic values in \cref{app:truncate}, including starting from an even larger prior, with a larger injected value of $m_1$ and lower $p_0$.

\begin{figure*}
    \centering
    \includegraphics[width=\textwidth]{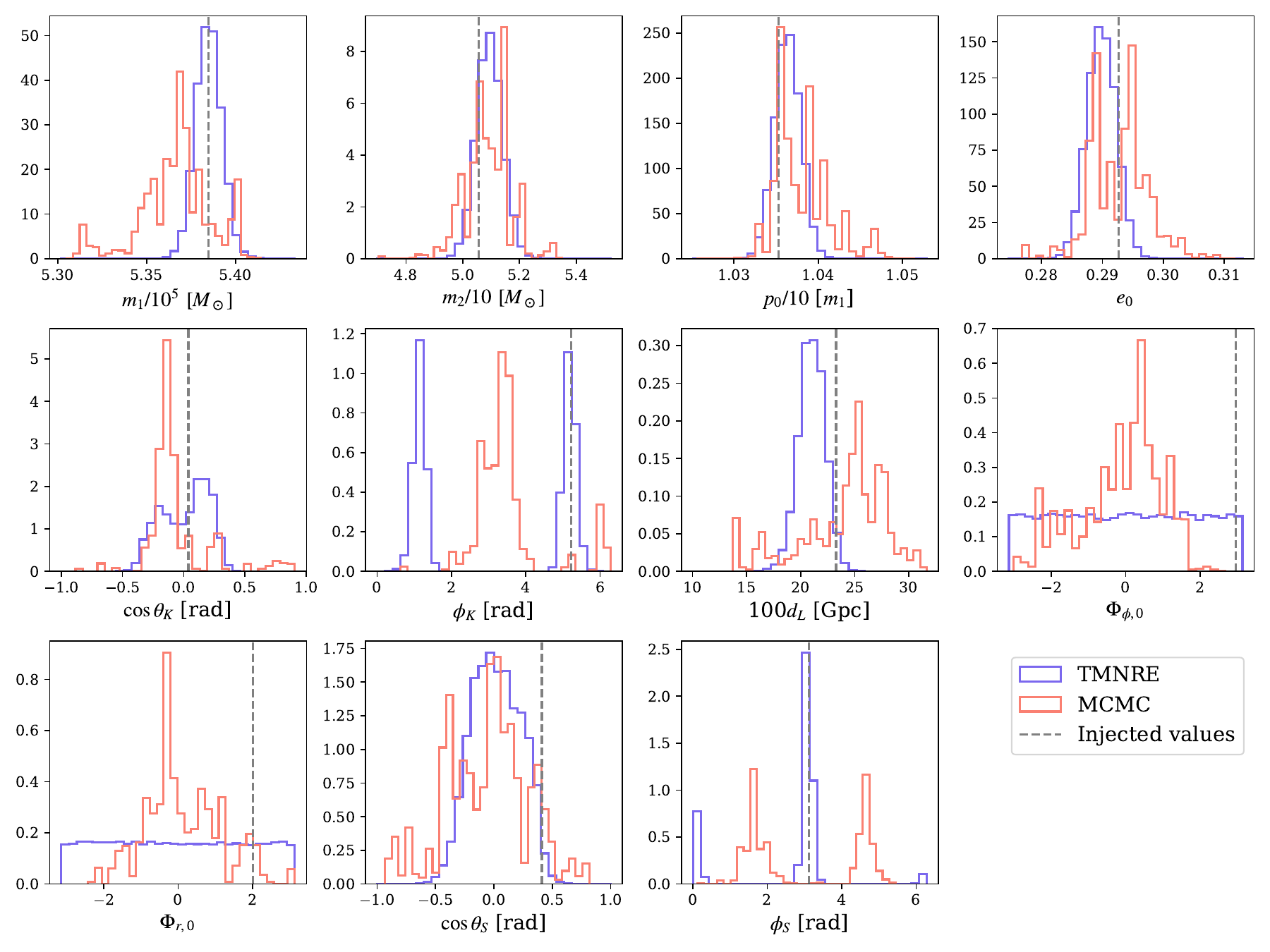}

    \caption{Comparison of TMNRE and MCMC approaches using the same priors identified by the truncation after the 5th round of inference by TMNRE. The blue distributions show the estimated proposal distributions from the 6th round of TMNRE, trained on 150000 simulations. The red distributions are computed from an MCMC run where the starting points are initialised randomly from the prior, with 32 walkers and 4687 steps. This conservatively sets the number of waveform evaluations to be approximately 150000 such that it is a fair comparison with the 6th round of TMNRE. On inspecting the MCMC chains, they are clearly far from convergence. The true injected values are shown by the grey dashed lines.
    }
    \label{fig:PE_comparison}
\end{figure*}

We now compare our results against those from a traditional MCMC approach in a regime where we can put the two methods on an equal footing in terms of their starting states and number of waveform evaluations. We run an MCMC as implemented in \texttt{eryn}~\cite{Karnesis:2023ras} with the affine-invariant proposal \texttt{StretchMove} \cite{2013PASP..125..306F,Goodman:2010dyf}, using the truncated priors from the sixth round of our TMNRE pipeline, with 32 walkers, 4 temperatures, and a step limit of 4,687, ensuring that the number of likelihood evaluations remains (conservatively) comparable to that of the final TMNRE round (150,000 simulations). We show in \cref{fig:PE_comparison} that with these much wider priors than are usually used for MCMC and with the same number of waveform evaluations as the TMNRE pipeline, the MCMC chains do not converge. For many of the parameters, TMNRE does significantly better at estimating the proposal distributions. This MCMC run took 49 hours on a GPU, compared to 14 hours for the final round of TMNRE. This comparison predominantly demonstrates the relative efficiency at which the TMNRE approach converges with respect to MCMC. Since optimising the MCMC approach is out of the scope of this work, we do not claim that it could not do better if given longer to thermalise.

\subsection{Additional results}

\begin{figure*}
    \centering
    \includegraphics[width=0.6\textwidth]{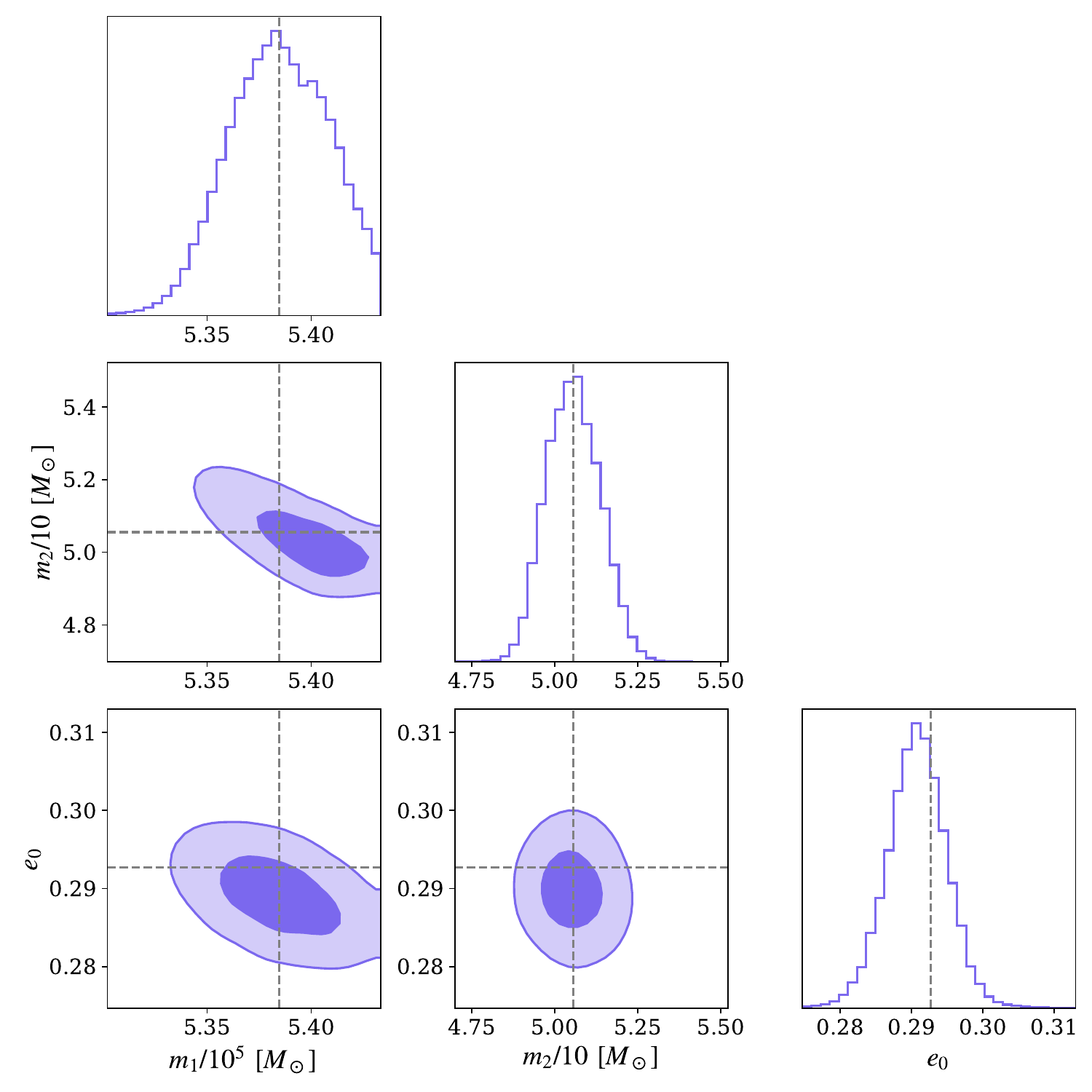}

    \caption{2d marginal proposal distributions for the black-hole masses and eccentricity, trained in parallel with the 1d marginals. The light shaded regions are the $2\sigma$ contours while the dark shaded regions are the $1\sigma$ contours. We used the same $1.5\times 10^5$ simulations from the 6th round of TMNRE that produced the 1d marginals in \cref{fig:PE_truncate}. The true injected values are shown by the grey dashed lines.} 
    \label{fig:2d_corner}
\end{figure*}

\noindent It is also possible to estimate the higher-dimensional marginal proposal distributions with this method. Since each $n$D marginal is trained independently, there is no guarantee that given a particular network architecture, the proposal distributions will be consistent with each other given that they are not lower-dimensional projections of a full joint posterior. However, having zoomed in on the parameter space with 6 rounds of truncation, we train the 2D marginals for the masses and eccentricities, and show in \cref{fig:2d_corner} that despite being trained independently of each other, they are consistent with having been projected from the full joint distribution. We can quantify this by checking the relative difference between the $2\sigma$ credible intervals for the 1d and 2d distributions, and find that for $m_1$ the difference is less than $50\,{\rm M_\odot}$ for each distribution, for $m_2$ it is less than $0.1\,{\rm M_\odot}$ and for $e_0$ it is less than 0.002. The 2D marginals were trained on the same set of simulations as were used to produce the 1D marginals in \cref{fig:PE_truncate}, and the training and validation losses are summed over all marginals.

However, we caution that our set-up appears to break down for some of the other parameters, especially those exhibiting multi-modality in their 1D marginals, resulting in 2D marginals that are much broader than their 1D counterparts. This may require different summaries to be shown to the $n$D logratio estimator, and we leave a full diagnosis of this issue to future work. We stress that the marginal aspect of this pipeline has many benefits in terms of simulation-efficiency when narrowing down the parameter space as the number of simulations required to reach a particular level of precision scales more favourably with the number of parameters than likelihood-based methods. This could also be extremely valuable when considering the more complicated LISA data analysis problem of dealing with many overlapping sources and non-stationary, non-Gaussian noise. Not only will the dimensionality of the problem increase dramatically, but there will also likely be many nuisance parameters that appear in the forward model. Marginal neural ratio estimation enables effortless marginalisation over nuisance parameters \cite{Miller:2021hys} that would otherwise be a bottleneck for many methods that estimate the full joint posterior.

\section{Conclusions and outlook}\label{sec:outlook}

\noindent We have explored the use of sequential simulation-based inference for search and parameter estimation of extreme mass ratio inspirals in future space-based gravitational wave detector data. We showed that TMNRE can be employed as a tool to very efficiently narrow down the search parameter space, meaning it could form a useful part of the search pipeline. In its current form, this method also estimates informative proposal distributions for all parameters, with the exception of the initial phases of the system.

We found that the proposals estimated by this set-up are much more informative than those achievable with MCMC, if compared over the same number of simulations or waveform evaluations with the MCMC chains initialised from the same final round priors as the TMNRE approach. However, we note that the extremely narrow posteriors that are in principle realisable for EMRIs, seen, for example, when MCMC chains are initialised very close to the true value, were not achieved in this TMNRE set-up. We believe this to be in part due to the truncation scheme saturating conservatively early in the presence of multi-modal proposal distributions, as well as the need for a more sophisticated data-compression scheme in order for the network to identify differences between waveforms that are dominated by noise. To expand on this point, we believe that the saturation in the sensitivity could be attributed to the dimensionality of the data input. In particular, in the current form, the network must learn subtle features of the signal vs. noise distribution in each data bin. For a large number of bins, this requires a very large number of data examples and potentially significantly larger network capacity, even if the information could be obtained from a significantly lower dimensional summary. This motivates exploring pre-conditioned data summaries for these long duration signals. This will also be vital for handling longer and more distant signals without compromising on sampling rate, which has been shown to be important for EMRI data analysis \cite{fastfourier}. Furthermore, a different parametrisation of the search space might reduce the multimodality of the problem and hence improve the accuracy of this setup. We also comment here on the other SBI options that are available beyond the TMNRE setup. For example, after using the current algorithm to efficiently narrow down the parameter space, the learned proposal distribution could be used directly in a density estimation (such as neural posterior estimation~\cite{2016arXiv160506376P}) or a higher-dimensional autoregressive estimation setup. This would allow us to maintain the benefits of simulation efficiency in the search phase, but obtain full information about the joint posterior in later stages of the pipeline. Additionally, the TMNRE proposal could be used to inform some form of data preconditioning (see e.g. Ref.~\cite{Dax:2024mcn}) to optimise the compression step that we argue is required to obtain full sensitivity.

On the signal modeling side, our current analysis focused on Schwarzschild EMRIs in vacuum, but the next step is to include higher-order effects, such as spin, which introduce additional complexities to the waveform. Given the flexibility of SBI methods, which only require simulation data, incorporating these effects should be straightforward in terms of implementation. We plan to test our framework on the upcoming version 2.0.0 of \texttt{FEW} that models Kerr EMRI systems  \cite{Katz:2021yft,Speri:2024qak}.

We have assumed Gaussian noise for simplicity in this work. However, real LISA data will exhibit non-stationary and non-Gaussian noise features due to instrument drifts \cite{Spadaro:2023muy}, gaps in data acquisition \cite{Burke:2025bun}, and overlapping signals. This aspect of the LISA data-analysis problem is where SBI could prove to be extremely powerful, since computing a likelihood for an EMRI signal that incorporates these noise complications is challenging. On the other hand, SBI solely requires a computational recipe to simulate the noise, which can then be fed directly into the neural network in the way we have described for the Gaussian noise in this work. We would like to explore the potential that this framework has for combatting this issue, as it could be uniquely useful in this EMRI regime.

Ultimately, the goal is to integrate EMRI data analysis into a broader LISA global fit~\cite{Littenberg:2023xpl,Katz:2024oqg,Strub:2024kbe,Deng:2025wgk}. This involves searching for and measuring the parameters of multiple populations of sources simultaneously. The SBI approach explored here has the potential to play a role in such a pipeline, either by providing refined proposal distributions for EMRIs after an initial search stage or by contributing directly to a hierarchical global fit.

More broadly, our pipeline can serve as a building block for further investigation into the use of sequential simulation-based inference for space-based gravitational wave data. Future improvements will incorporate increasing complexity on both the signal and noise modeling sides, as well as dealing with the fact that EMRIs are just one species of the broader population of systems that will be present in the data.

\acknowledgements
\noindent P.C. and D.G. are supported by ERC Starting Grant No.~945155--GWmining, Cariplo Foundation Grant No.~2021-0555, MUR PRIN Grant No.~2022-Z9X4XS, MUR Grant ``Progetto Dipartimenti di Eccellenza 2023-2027'' (BiCoQ), and the ICSC National Research Centre funded by NextGenerationEU. D.G. is supported by MSCA Fellowship No. 101064542–StochRewind, MSCA Fellowship No. 101149270–ProtoBH, and MUR Young Researchers Grant No. SOE2024-0000125. Computational work was performed at CINECA with allocations through INFN, Bicocca, and an ISCRA Class C project grant. J.A. is supported by a fellowship from the Kavli Foundation. This work is part of a project that has received funding from the European Research Council (ERC) under the European Union’s Horizon 2020 research and innovation programme (Grant agreement No. 864035).

\bibliographystyle{apsrev4-1}
\bibliography{sbiemri}

\appendix
 
\section{Convergence tests}\label{app:pp}
\begin{figure*}
    \centering
    \includegraphics[width=\textwidth]{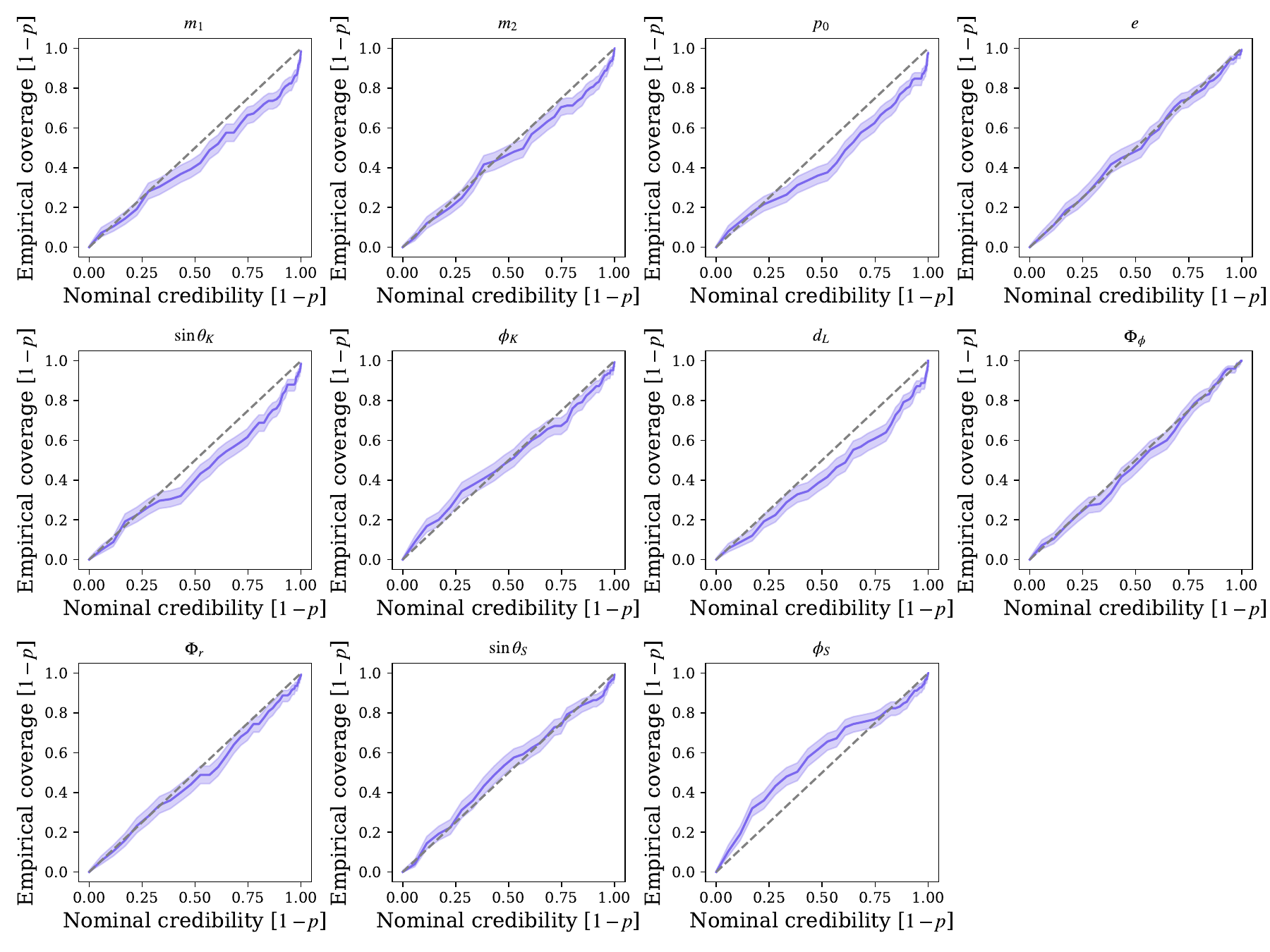}

    \caption{$p-p$ plots for each of the 11 Schwarzschild EMRI parameters with 500 samples drawn from the 6th round priors of the TMNRE pipeline. The dashed grey diagonal lines represent perfect agreement.}
    \label{fig:pp_truncate}
\end{figure*}

\noindent $p-p$ plots for each parameter are shown in \cref{fig:pp_truncate} and are calculated with 500 samples from the 6th-round priors, again with the marginals trained in parallel. The empirical coverage is an estimate of the probability mass of the learned proposal distributions that lies within the credible region around the true value. If the empirical coverage as a function of the nominal credibility lies above (below) the diagonal, then the estimated proposals are conservatively wide (optimistically narrow). Our $p-p$ plots are (close to) diagonal and demonstrate reasonable coverage across all the parameters. There is some evidence of overconfidence for $m_1$, $p_0$, $d_L$ and $\theta_K$, and some evidence of underconfidence for $\phi_S$. We expect that this is partly due to the fact that the marginals are trained in parallel, with the training and validation losses summed across the parameters. This means that individual parameters for which the network has converged could be overfit if another parameter is dominating the continued decrease in the losses.

\begin{figure*}
    \includegraphics[width=0.5\textwidth]{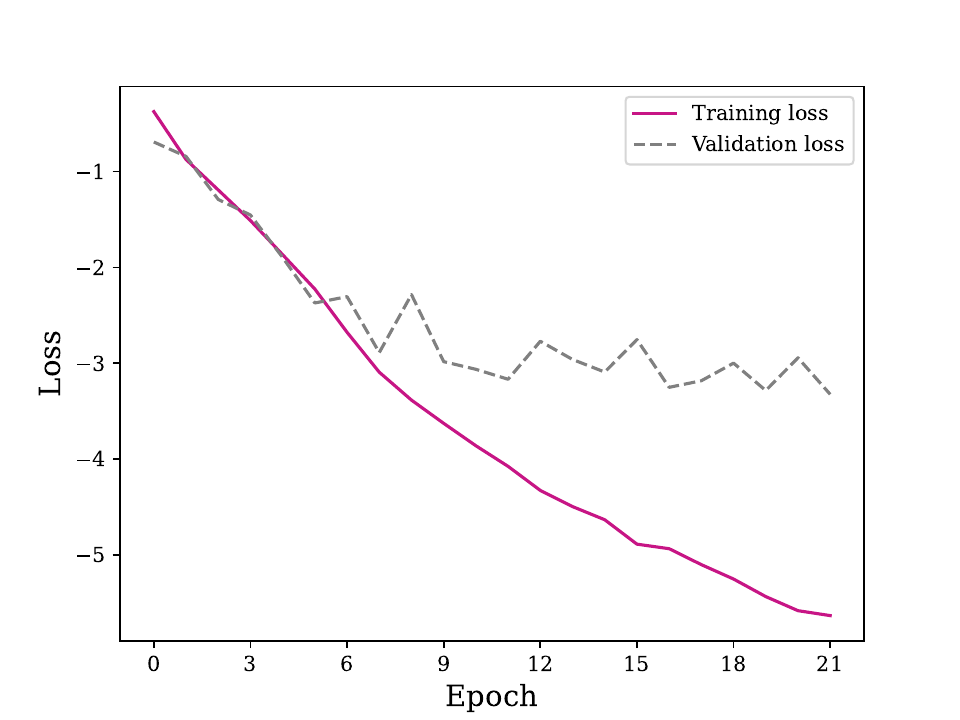}

    \caption{The training and validation losses of the network from the sixth round of TMNRE which produced the results for the fiducial system in this paper, namely the proposal distributions and truncation bounds in \cref{fig:PE_truncate,fig:bound_compare}. The training loss (pink solid) is computed on-the-fly and here we show the value at the end of the epoch, while the validation loss (grey dashed) is computed just once at the end of each epoch.}
    \label{fig:trainval}
\end{figure*}

We also show in \cref{fig:trainval} the values of the training and validation losses at the end of each epoch for the network which produced the results in \cref{fig:PE_truncate,fig:bound_compare}. In this sixth round, the network trained for 28 epochs, with no decrease in the validation loss for the final 7 epochs (equal to the early stopping criterion), and hence the weights of the network were returned to the state after 21 epochs. There is some evidence of overfitting in the latter stages of training, which could also explain the overconfidence observed in \cref{fig:pp_truncate}.

\section{Truncation for intrinsic parameters}\label{app:truncate}

\noindent Here we discuss some additional examples of narrowing down the parameter space for intrinsic parameters. When we draw new values of $m_1$, $m_2$ and $p_0$ individually, keeping all other parameters fixed to the fiducial values in the main body of the paper, we find that the ability to narrow down the parameter space degrades considerably with $V_i/V_f=\{5.0, 55, 2.9\}$ for $m_1=8.3\times10^5\,{\rm M_\odot}$, $m_2=50.6{\rm M_\odot}$ and $p_0/m_1 = 13.0$ respectively. This is expected, since for these injected values, the signals are from much earlier phases of the inspiral instead of the final part before plunge where most of the information about the parameters is contained. For reference, the SNR of this system is $\sim70$.

\begin{figure*}
    \centering
    \includegraphics[width=\textwidth]{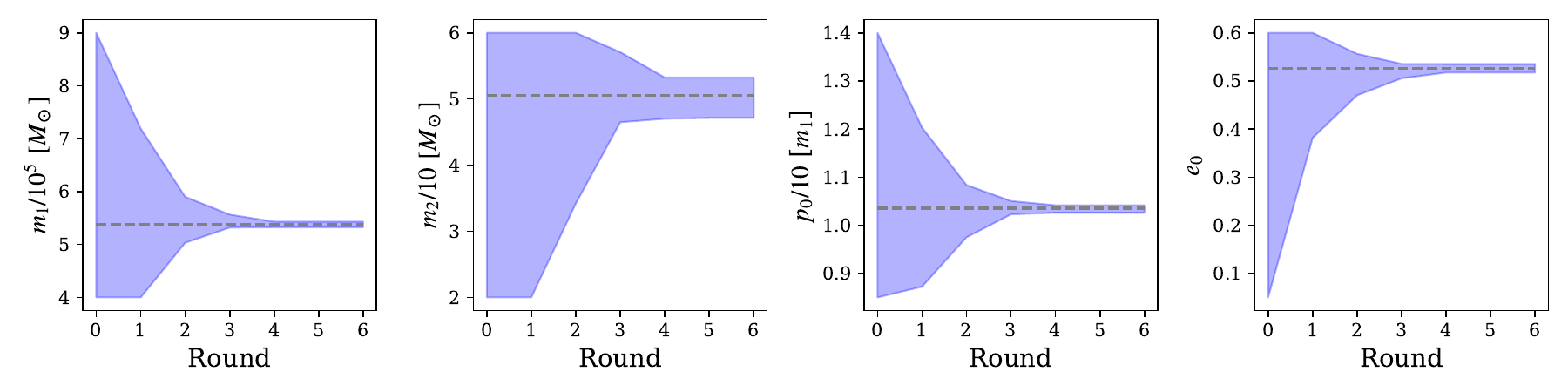}

    \caption{Shaded regions show the parameter space that is identified by TMNRE as still containing significant likelihood-to-evidence ratio weight for the four intrinsic parameters where $e_0=0.53$ is different to that in the main body of the paper, in 6 sequential rounds, trained on 150000 simulations in each round. The initial priors are shown by the shaded region at round 0, and the grey lines show the injected parameters. The total prior volume for all 11 parameters is narrowed down by a factor of $V_i/V_f = 1.1\times10^6$ after 6 sequential rounds of TMNRE.}
    \label{fig:e0_truncate}
\end{figure*}

\begin{figure*}
    \centering
    \includegraphics[width=\textwidth]{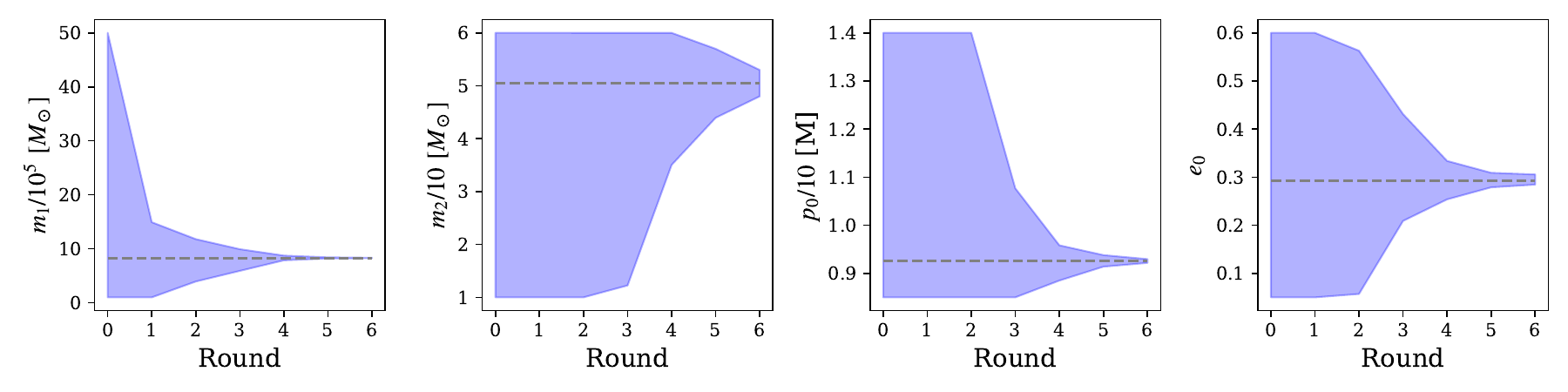}

    \caption{Shaded regions show the parameter space that is identified by TMNRE as still containing significant likelihood-to-evidence ratio weight for the four intrinsic parameters, with $m_1=8.3\times10^5\,{\rm M_\odot}$ and $p_0/m_1=9.25$ shifted with respect to the system in the main body of the paper, in 6 sequential rounds, trained on 150000 simulations in each round. The initial priors are shown by the shaded region at round 0 and are also wider than those used in the main body of the paper, and the grey lines show the injected parameters. The total prior volume for all 11 parameters is narrowed down by a factor of $V_i/V_f = 6\times10^7$ after 6 sequential rounds of TMNRE.}
    \label{fig:truncate_wideprior}
\end{figure*}

However for a new injected value of the orbital eccentricity, it is again the final part of the signal which is shown to the network and for $e_0=0.53$ we find $V_i/V_f=1.1\times10^6$ with truncation bounds shown in \cref{fig:e0_truncate}. Finally, we show in \cref{fig:truncate_wideprior} the truncation starting from even larger priors and for a larger value of $m_1=8.3\times10^5\,{\rm M_\odot}$ and lower value of $p_0/m_1=9.25$, such that the injection again captures the full final tenth of a year. The total prior volume for all 11 parameters is narrowed down by a factor of $V_i/V_f = 6\times10^7$ after 6 sequential rounds of TMNRE. The SNR of this system is 530.

\clearpage

\end{document}